\documentclass[11pt]{article}
\pdfoutput=1
\usepackage{graphicx,amssymb,amsfonts,amsmath,amssymb,amscd,amstext, mathrsfs}
\usepackage{color}
\makeatletter

\@addtoreset{equation}{section}
\newcommand{\be}{\begin{eqnarray}}
\newcommand{\ee}{\end{eqnarray}}
\newcommand{\ba}{\begin{array}}
\newcommand{\ea}{\end{array}}
\newcommand{\nn}{\nonumber}

\makeatletter \@addtoreset{equation}{section} \makeatother

\begin{document}
~
\vspace{0.5cm}
\begin{center} {\Large \bf   Butterfly Velocity in Quadratic Gravity}\\ 
                                                  
\vspace{1cm}

                      Wung-Hong Huang \\
                      \vspace{.2cm}
                      Department of Physics\\
                       National Cheng Kung University\\
                       Tainan, Taiwan\\

\end{center}
\vspace{1cm}
\begin{center} 
{\large \bf  Abstract} 
\end{center}
We present a systematic procedure of  finding  the shock wave equation in  anisotropic spacetime of quadratic gravity with Lagrangian ${\cal L}=R+ \Lambda+\alpha R_{\mu\nu\sigma\rho}R^{\mu\nu\sigma\rho}+\beta R_{\mu\nu}R^{\mu\nu}+\gamma R^2+{\cal L}_{\rm matter}$.  The general  formula of  the butterfly velocity is derived. We show that  the shock wave equation in the planar, spherical or hyperbolic black hole spacetime  of Einstein-Gauss-Bonnet gravity  is the same as that in Einstein gravity if  space is isotropic.    We consider the   modified AdS spacetime deformed by the  leading correction of  the quadratic curvatures and find that the fourth order derivative shock wave equation  leads to  two butterfly velocities if  $4\alpha+\beta<0$.   We also show that  the butterfly velocity in the D=4  planar black hole  is not   corrected by the quadratic gravity if $ 4\alpha+\beta=0$, which includes the $ R^2$ gravity.    In general, the correction of  butterfly velocity by the quadratic gravity may be positive or  negative, depends on the values of  $\alpha$, $\beta$,   $\gamma$   and temperature.   We also investigate the butterfly velocity in the Gauss-Bonnet massive  gravity.
\\
\\
\\
\\
\\
\\
\\
\\
whhwung@mail.ncku.edu.tw

\newpage
\tableofcontents
\newpage
~\\
\section {Introduction}

The notion of the out-of-time-order (OTO) correlator was introduced  in the context of semi-classical methods in superconductivity many years ago \cite{Larkin1969}. 
Recently, it has received renewed attention in the context of the AdS/CFT correspondence, where the butterfly effects (quantum chaos) can be studied by looking at black holes. 
Many novel aspects and interesting properties have been found \cite{Shenker1306, Shenker1312, Leichenauer1405,  Roberts1409, Roberts1412, Shenker1412, Maldacena1503, Fitzpatrick1601, Roberts1603}. 
The OTO four-point function between a pair of local operators $W(t,x)$, $V(t,x)$ is defined by 
\be
\label{def}
\langle [W(t,x),V(0)]^2\rangle_T \sim e^{\lambda(t-t_*-{|x|\over v_B})} \ .
\ee 
For the case of $W(t)=x(t)$ and $V(0)=p_x$, in the semiclassical limit where $[x(x),p_x] =i\hbar\{x(t),p_x\} =i\hbar{\partial x(t)\over \partial x(0)}$ which describes how the final position depends on the small changes of the initial position, this correlator can be used to diagnose the quantum chaos.  
The constant $t_*$ in \eqref{def} is the scrambling time at which the commutator grows to be ${\cal O}(1)$. 
The buttery velocity $v_B$ characterizes speed at which the perturbation grows; the Lyapunov exponent $\lambda$ measures the rate of growth of chaos.

It was found that the Lyapunov exponent is bounded by temperature T, $\lambda \le 2\pi\beta$ where $\beta={1\over T}$ \cite{Maldacena1503}. 
The inequality saturates for a thermal system that has a dual black hole described by the Einstein gravity.   The quantum correction to the chaos, which modifies the Lyapunov was studied in \cite{Fitzpatrick1601}.  Butterfly velocity has also an interesting property that it is  related to the diffusion constant \cite{Blake1603, Blake1604, Blake1611,Blake1705}. 

In the context of holography, the  shock wave near the black hole horizon provides a description of the butterﬂy effect in the dual field theory and butterfly velocity is identified by the velocity of shock wave, which describes how the perturbation spreads in space \cite{Shenker1306, Shenker1312}.  The method of finding the shock wave velocity for the general spacetime with matters was described in many years ago \cite{t'Hooft1985, Sfetsos9408}.  
Such a method has been used to obtain the butterfly velocity in the dual field theory by studying shock waves of the various black holes \cite{Reynolds1604, Sircar1602,  Huang1609, Feng1701, Cai1704, Jahnke1708, Mezei1608, Alishahiha1610, Caceres1512, Qaemmaqami1705, Qaemmaqami1707, Li1707, Ling1610,  Ling1610a,  Giataganas1708, Ahn1708,Huang1710, Li1710, Peng1802}.  
In \cite{Shenker1306}, it was found that the speed of propagation is
\be
v_B=\sqrt{D-1\over 2(D-2)} \ ,
\ee
where $D$ is the spacetime dimension of the bulk theory. 
Later investigations have extended this result to anisotropic black holes/branes \cite{ Jahnke1708,  Qaemmaqami1705,  Qaemmaqami1707, Ling1610, Giataganas1708,  Huang1710}. 

 The butterfly velocities described by the higher derivative gravities, including 3D massive gravity and Gauss-Bonnet gravity, were discussed in \cite{Roberts1409, Mezei1608, Alishahiha1610,   Qaemmaqami1705,  Qaemmaqami1707, Huang1710}.  It was found that the Gauss-Bonnet term (in the absence of matters) modifies the screaming time but does not change the shock wave equation \cite{Roberts1409}.\footnote{The Gauss-Bonnet gravity is the simplest quadratic correction of the Einstein theory without introducing derivatives higher than second in the field equation.  Higher derivative corrections arise due to stringy corrections of the classical action. 
In terms of AdS/CFT correspondence this corresponds to next-to-leading order corrections in the 1/N expansion of the dual CFT \cite{Gubser9805, Nojiri0006}.}   The butterfly velocity in four-derivative gravity for isotropic spacetimes was partially studied in \cite{Mezei1608}.

Higher-derivative gravities find many applications in the study of holography.
For instance, the authors of \cite {Kats0712, Brigante0712, Brigante0802, Myers0812} considered a Lagrangian 
\be
\label{L}
{\cal L}=R+ \Lambda + \alpha R_{abcd}R^{abcd}+\beta R_{ab}R^{ab}+\gamma R^2 +{\cal L}_{\rm matter}
\ee
and found that for a small value of quadratic curvature the ratio of the shear viscosity $\eta$ to the entropy density $s$ is modified by ${\eta\over s}={1\over 4\pi}(1-4 \alpha)$.  
The viscosity bound established in \cite{Kovtun0309} is therefore violated in higher derivative gravities for $ \alpha>0$.   
Holographic superconductors with higher curvature corrections had  been extensively studied  \cite {Gregory0907, Barclay1009, Pan0912}.  Holographic shear sum rule in Einstein gravity corrected by squared curvature was studied in \cite{Chowdhury1711}.  The higher derivative terms are shown to have a strong impact on the bound on charge diffusion \cite{Baggioli1612, Baggioli1705}.

In the holography  the addition of higher derivative corrections to gravity theories could be used to probe the dual physics moving away from infinite N. For example, in the study of  viscosity bound violation Kats and Petrov \cite{Kats0712} showed that  $\alpha={1\over 2{\rm N}}$ for $N\gg1$.  

 Note  that at leading order of $\alpha, \, \beta,\, \gamma$  only $\alpha$ is unambiguous while $ \beta$ and $\gamma$ can be arbitrarily altered by a field redefinition \cite{Kats0712}.  However, the coefficients  $ \beta$ and $\gamma$ in \eqref{L} are physical once the matter fields are turned on. For example, in an Einstein-Maxwell theory, shifting away the coefficients  $ \beta$ and $\gamma$ will generate new mixed terms of the form $RF^2$ and $R_{\mu\nu}F^{\mu\lambda}F^\nu_\lambda$, which is relevant in the studies of R-charged backgrounds \cite{Cremonini0910}.
 
   Note also  that in the context of string theory and holography the higher-order gravity are viewed as part of infinite series of  corrections to the leading order string effective action. The action of  higher-order gravity is not quantized, which makes the issue of ghostlike modes moot.   Therefore, in despite of the existence of ghosts in the higher-order gravity except some special cases, such as Gauss-Bonnet gravity,  we will study in this paper the general case with three coefficients of   $\alpha, \, \beta,\, \gamma$.  Since that the D+1 dimensional  bulk theory will dual to D-dimensional boundary field theory the investigation in this paper can then be applied to any D-dimensional field theory.

In the previous paper \cite{Huang1710}, we had considered the Gauss-Bonnet gravity with arbitrary matter fields and derived a general formula of butterfly velocity.  
We calculated the butterfly velocity in planar, spherical, and hyperbolic black holes for the Gauss-Bonnet gravity, with Maxwell and scalar fields.
The goal of this work is to consider butterfly velocity in anisotropic space of more general quadratic gravity \eqref{L} with arbitrary matters.  

The paper is organized as follows.

In section 2, we briefly discuss some basic properties of quadratic gravity, including the $R^2$ gravity, Gauss-Bonnet gravity,  conformal gravity and Gauss-Bonnet massive gravity. 

In section 3, after describing general anisotropic space, we briefly discuss the Kruskal coordinates and derivation of  the shock wave equation.  The shock wave equation of  Einstein gravity is presented in \eqref{Einstein}.  The derivations were  detailed in our previous paper.

In section 4, we first collect five relations and then use them to simplify the shock wave equation in the quadratic gravity. The equation is shown  in \eqref{totalL3} which has only six  terms.  Then,  after  calculations  the exact formulas of the six term are presented.  Using the formulas, as a simple example we obtain the formula of butterfly velocity in the anisotropic space of Gauss-Bonnet gravity theory. The formula is shown  in \eqref{mainresult2} which, in the case of isotropic space,  reproduces the equation (4.18)  in our previous paper \cite{Huang1710}. The double summations in  the formula \eqref{mainresult2} means that the shock wave equation in the anisotropic space is a non-trivial extension of that in  the  isotropic space. Using the formula we prove that :  {\it In the D-dimensional planar, spherical or hyperbolic black hole spacetime  the Einstein-Gauss-Bonnet gravity has the same shock wave equation as that in Einstein gravity if  and only if the space is isotropic}. 
  
   In section 5 we  simplify  the formula to case of isotropic spacetime in which the black hole is planar, spherical or hyperbolic.   Then we apply the result to  the metric in \cite{Kats0712}, which is an analytic solution in leading order of quadratic gravity,  to calculate the butterfly velocity.   The final  formulas are shown in eq.\eqref{finalVB1} and eq.\eqref{finalVB2}.   In  higher derivative gravity the shock wave equation is  fourth order and the metric provides two sources for two operators in the context of AdS/CFT correspondence.   Since that each operators results to a butterfly velocity and in general we have two butterfly velocities \cite{Mezei1608, Alishahiha1610}. However, we show that : {\it Only if $ 4\alpha+\beta<0$ could the second velocity  appear}.  We also prove that : {\it   The butterfly velocity in D=4  planar black hole  does not be  corrected by the quadratic gravity if $ \beta +4\alpha=0$, which includes the $ R^2$ gravity}.   We present various explicit  butterfly velocity in the Gauss-Bonnet gravity, conformal gravity,  and $R^2$ gravity respectively. We  see that, depending on the values of $\alpha$, $\beta$ and $\gamma$ the velocity correction from the quadratic gravity may be from positive to negative or from negative to positive while increasing the temperature.  We use our formula to  check the two butterfly velocities in a special quadratic gravity which was first derived in Alishahiha's paper \cite{Alishahiha1610}. The butterfly velocity in the Gauss-Bonnet massive  gravity is also studied.
   
    Last section is a short summary.  Many detailed tensor calculations are collected in appendix.

\section {Quadratic Gravity and Gauss-Bonnet Massive  Gravity}
\subsection {Quadratic Gravity}
We are interested in the general higher-derivative gravity described in \eqref{L}.  The generalized gravitational equation is
\be
{\sf G}_{ab}&=&T_{ab} \ ,
\ee
where the generalized Einstein tensor is defined by
\be
\label{get}
{\sf G}_{ab}&=&R_{ab}+K_{ab}+D_{ab}-{1\over 2} g_{ab} \Big({\cal L}-(\beta+4\gamma)\Box R \Big) \ ,
\ee
with
\be
K_{ab}&=&2\alpha R_{acde}R_b^{~cde} +2(2\alpha+\beta)R_{acbd}R^{cd}-4\alpha R_{ac}R_b^{c}+2\gamma R_{ab}R\ , \\
D_{ab}&=&(4\alpha+\beta)\Box R_{ab}-(2\alpha+\beta+2\gamma)\nabla_a\nabla_b R \ .
\ee
Note that $D_{ab}=0$ when $\beta=-4\alpha$, $\alpha=-\gamma$:
this is the Gauss-Bonnet gravity where terms with higher derivatives (more than second-derivative) do not appear. 

The Gauss-Bonnet gravity is non-renormalizable but ghost free.  It can be shown that  the inclusion of the GB term entails causality violation \cite{Camanho1407}\footnote{We are grateful to J. D. Edelstein for pointing it out.}.  However, this does not totally rule out its uses in the holographic context. In the case $\alpha={-\beta\over2}={\gamma\over3}$, one has conformal/Weyl gravity which has ghosts, though it is renormalizable \cite{Stelle1977, Adler1982}.\footnote{It was discussed in \cite{Maldacena1105, Anastasiou1608} that Einstein gravity may emerge from conformal gravity upon imposing suitable boundary conditions eliminating ghosts.}  The $R^2$ gravity is a simple modification to the Einstein gravity with various interesting applications. One can, for instance, include $R^2$ to study the late-time expansion of the cosmic acceleration \cite{Sotiriou0805}.  We will consider butterfly velocities in all these cases.  More discussions about the higher-derivative gravities can be  found in \cite{Bueno1610}. 
\subsection {Gauss-Bonnet Massive  Gravity}
In recent a new class of nontrivial massive black holes in AdS spacetime was studied in \cite{Vegh1301, Blake1308, Cai1409}. In the theory of the massive gravity, the optical conductivity shows an emergent scaling law which is consistent with that found earlier by Horowitz, Santos, and Tong \cite{Horowitz1204} who introduced an explicit inhomogeneous lattice into the system.

 Note that the mass terms of the gravitons  will be plagued by various instabilities sometimes at the non-linear level. The authors of \cite{Rham1011, Hassan1106, Hassan1109} constructed a theory where the Boulware-Deser ghost \cite{Boulware1972} was eliminated by introducing higher order interaction terms into the Lagrangian which  is
 \be
 \label{MG}
 {\cal L}&=&R+\Lambda+m^2\sum_1^4c_i\,{\cal U}_i(g,f)+{\cal L}_{\rm matters}\\
 {\cal U}_1&=&[{\cal K}],~~~{\cal U}_2=[{\cal K}]^2-[{\cal K}^2],~~~{\cal U}_3=[{\cal K}]^3-3[{\cal K}][{\cal K}^2]+2[{\cal K}^3]\\
 {\cal U}_4&=&[{\cal K}]^4-6[{\cal K}]^2[{\cal K}^2]+8[{\cal K}][{\cal K}^3]+3[{\cal K}^2]^2-6[{\cal K}^4]
 \ee
 where $c_i$ are constants, and ${\cal K}^\mu_\nu=\sqrt{g^{\mu\alpha}f_{\alpha\nu}}$ with $[{\cal K}]={\cal K}^\mu_\mu$ for a reference metric $f_{\mu\nu}$. For a recent review of massive gravity in this context, see \cite{Hinterbichler1105}. 
 
   In the papers \cite{Hendi1507,Sadeghi1507, Parvizi 1704, Hendi1510, Hendi1708} the thermodynamics of black hole in Gauss-Bonnet massive gravity theory were studied and in this paper, for the sake of completeness, we will adopt the metric  solutions therein to  calculate the associated butterfly velocity.
\section {Kruskal Coordinate and Shock Wave Equation}
For self-consistent we briefly describe the Kruskal coordinate and sketch  the  shock wave equation.
\subsection{Kruskal Coordinate}
We will derive the formula of butterfly velocity in the following anisotropic background:
\be
ds^2&=&-a(r)f(r)dt^2+{dr^2\over b(r)f(r)}+\sum_{S=1}^n h^{(S)}(r)\,\bar g^{(S)}_{ij}(x)\,dx_{(S)}^idx_{(S)}^j
\ee
The horizon locates at $r=r_H$ then $f(r_H)=0$ while $a(r_H)\ne0$ and $b(r_H)\ne0$.  
The associated temperature of  black hole or black brane is 
\be
T={f\rq{}(r_H)\sqrt {a(r_H)b(r_H)}\over 4\pi}
\ee
Defining the tortoise coordinate $r_*$  the line element of time and radial parts  can be expressed as
\be
d\bar s^2&=&-a(r)f(r)dt^2+{dr^2\over b(r)f(r)}=-a(r)f(r)\Big[dt^2-dr_*^2\Big]\\
dr_*&=&{dr\over f(r)\sqrt{a(r)b(r)}}
\ee
The metric can be written in Kruskal coordinate 
\be
\label{metric0}
ds^2&=&2A(UV)dUdV+\sum_Sh^{(S)}(UV)\,\bar g^{(S)}_{ij}(x)dx_{(S)}^idx_{(S)}^j
\ee
where
\be
A(UV)&=&-{2a(r)f(r)\over f\rq{}(r_H)^2a(r_H)b(r_H)}e^{-\sqrt{a(r_H)b(r_H)}~f\rq{}(r_H)~r_*}\\
U&=&e^{\sqrt{a(r_H)b(r_H)}{f\rq{}(r_H)\over 2} (-t+r_*)}\\
V&=&e^{\sqrt{a(r_H)b(r_H)}{f\rq{}(r_H)\over 2} (t+r_*)}\\
r_*&=&{1\over \sqrt{a(r_H)b(r_H)}f\rq{}(r_H)}\ln (UV)
\ee
In the tortoise coordinate $r_*(r_H)=-\infty$ and thus on the horizon $U_H=0$.   Above definitions  imply following useful relations
\be
\label{A1}
 A(U_H)&=&{2r_H\over f\rq{}(r_H)b(r_H)}\\
\label{h1}
h'(U_H)&=&{dh(UV)\over dr}{dr\over d(UV)}\mid_{r=r_H}=r_H~h'(r_H)\\
A'(U_H)&=&\left({dA(UV)\over dr}{dr\over d(UV)} \right)_  {r=r_H}=
{r_H^2\over b(r_H)f'(r_H)}~\Big( {3a'(r_H) \over a(r_H) }+{b'(r_H) \over b(r_H) }+{2f''(r_H)\over f'(r_H)}\Big)~~
\ee
Higher derivative terms $h''(U_H)$ and $A''(U_H)$ can be evaluated in the similar way. 
\subsection {Shock Wave Equation}
In the Kruskal coordinate the  generalized gravitational  equation can be expressed  as 
\be
{\sf G}=T_{matter}&=&2T_{UV}(U,V,x)dUdV+T_{UU}(U,V,x)dUdU+T_{VV}(U,V,x)dVdV \nn\\
&&+\sum_ST^{(S)}_{ij}(U,V,x)dx_{(S)}^i dx_{(S)}^j
\ee 
Along the arguments of Dray and G. t'Hooft \cite{t'Hooft1985}, after adding a small null perturbation of asymptotic energy $E$ 
\be
T_{(shock)\hat U\hat U}=E\,e^{2\pi t/\beta}~a(\hat x)~\delta(\hat U)
\ee 
the spacetime is still described by \eqref{metric0} but $V$ is shifted by 
\be
V \rightarrow V + \alpha(x)
\ee
Through analysis we can find that in terms of  the new coordinates \cite{Sfetsos9408}
\be
\hat U=U,~~~\hat V=V+\Theta(U) \alpha(x)
\ee
where $\Theta=\Theta(U)$ is a step function, the metric can be expressed by 
\be
\label{metric1}
ds^2&=&2\hat A(\hat U,\hat V)d\hat Ud\hat V+\sum_S\hat g^{(S)}_{ij}(\hat U,\hat V,\hat x) d\hat  x_{(S)}^id\hat x_{(S)}^j - 2\hat A~\hat \alpha(\hat x)\hat \delta(\hat U) d\hat U^2
\ee
the generalized gravitational  equation, after dropping  hat notation,  becomes 
\be
\label{sweq}
{\sf G}^{(1)}_{UU}+2{\sf G}^{(0)}_{UV}~ \alpha(x) ~\delta(U)=E\,e^{2\pi t/\beta}~a(x)~\delta(U)
\ee
Term ${\sf G}^{(1)}_{UU}$ and ${\sf G}^{(0)}_{UV}$ are the first-order correction and zero-order generalized Einstein tensor in the metric  \eqref{metric1} respectively.  This is  the shock wave equation.   Using above formulation we will present a systematic procedure to  find the  differential equation of  $ \alpha(x)$ for the quadratic gravity in the anisotropic spacetime \eqref{metric1} and then obtain the associated formula of butterfly velocity.
\subsection{Shock Wave Equation in Einstein Gravity}
For Einstein gravity theory the tensor calculation in our previous paper  \cite{Huang1710} gives
\be
\label{Einstein}
{\sf G}^{(1)}_{UU}+2{\sf G}^{(0)}_{UV}~ \alpha(x) ~\delta(U)
&=&{A\over 2}\sum_S \Big({2\over h^{(S)}}~\bar\Delta^{(S)}\alpha(x)- {d^{(S)}\,h'^{(S)}\over A h^{(S)}}\alpha(x)\Big)\,\delta(U)
\ee
and, in the  case with local source $a(x)=\delta(x_i^{(Q)})$, the shock wave equation becomes
\be
\label{mainresult1}
\Big[\bar\Delta^{(Q)}-h^{(Q)}(U_H)\sum_S {d^{(S)}\,h'^{(S)}(U_H)\over 2A(U_H)~ h^{(S)}(U_H)}\Big] ~\alpha(t,x_i^{(Q)}) = E\,e^{2\pi t/\beta}~{h^{(Q)}(U_H)\over A(U_H)}~\delta(x_i^{(Q)})
\ee
The butterfly velocity along the direct $x_i^{(Q)}$ is 
\be
\label{mainvelocity}
v_B^{(Q)}&=&{2\pi kT\over M_{(Q)}},~~
M^2_{(Q)}=h^{(Q)}(r_H)\sum_S {d^{(S)}}~{b(r_H) f'(r_H)h'^{(S)}(r_H)\over 4 h^{(S)}(r_H)}
\ee
where we have used the relations of \eqref{A1} and \eqref{h1}.   Formulas \eqref{Einstein} and \eqref{mainresult1} were  derived by us in  eq.(2.28) and eq.(2.30) in \cite{Huang1710}, respectively. 
\section{Shock Wave Equation in Quadratic Gravity}
To investigate the theory with quadratic gravity  we  first collect following primary relations which can be proved with the help of appendix A, B and C.  Note that we denote coordinate index  by $a,b,c,d$ and these not $U,V$ by $i,j,k,m,n$. 
\subsection{Five Relations}
Relation 1 : On the horizon the non-zero values of $R^a_{~bcd}$ are 
\be
R^U_{~UUV}&=&-R^U_{~UVU}\dot=-{A'(0)\over A(0)},~~~
R^{U(S)}_{~iUj}=-R^{U(S)}_{~ijU}\dot=-{\bar g^{(S)}_{ij}(x)h'^{(S)}(0)\over 2A(0)}\\
R^V_{~VUV}&=&-R^V_{~VVU}\dot={A'(0)\over A(0)},~~~R^{V(S)}_{~iVj}=-R^V_{~ijV}\dot=-{\bar g^{(S)}_{ij}(x)h'^{(S)}(0)\over 2A(0)}\\
R^{i(S)}_{~UVj}&=&-R^{i(S)}_{~UjV}\,=\,
R^{i(S)}_{~VUj}=-R^{i(S)}_{VjU} \,\dot=\,{h'^{(S)}(0)\over 2h^{(S)}(0)},~~
R^{i(S)}_{~jkm}=\bar R^{i(S)}_{~jkm}\,\dot\ne\,0
\ee

Relation 2 : On the horizon the non-zero values of $R_{ab}$ and $R$ are 
\be
R_{UV}&=&R_{VU}\dot=-{A'(0)\over A(0)}-\sum_S{{d^{(S)}}\,h'^{(S)}(0)\over 2h^{(S)}(0)}\\
R^{(S)}_{ij}&\dot=&\bar R^{(S)}_{ij}-{\bar g^{(S)}_{ij}(x)h'^{(S)}(0)\over A(0)}\\
R&\dot=&-{2A'(0)\over A(0)^2}+\sum_S {\bar R^{(S)}\over h^{(S)}(0)}-{2d^{(S)}\,h'^{(S)}(0)\over A(0)\,h^{(S)}(0)}
\ee
where the superscript  $(S)$  is used to specify the coordinate $dx^i_{(S)}$ in metric \eqref{metric0}. The notation $\dot=0$ is used to emphasize that  the value is calculated on the horizon.  Notice that the index $(S)$ in $R^{i(S)}_{~jkm}$ means that indices $i,j,k,m$ have to be on  the same $(S)$ otherwise the tensor is zero. 

 We use $\bar R^{i(S)}_{~jkm}$ ,  $\bar R^{(S)}_{ij}$ and  $\bar R^{(S)}$  to denote the curvature evaluated in  metric $ds^2=\bar g^{(S)}_{ij}(x) dx^idx^j$. Note that $d^{(S)}= {\bar g^{ij(S)}} \bar g^{(S)}_{ij}$ is the dimension of space $dx_{(S)}^i$ in \eqref{metric0}. The relation between bulk  dimension $D$ and dimension $d^{(S)}$ is 
\be
D=2+\sum_Sd^{(S)}
\ee 
In isotropic space above relation reduces to $D=2+d$ while other literature denotes  $D=1+d$. Our notation is convenient when space  is  anisotropic.

Relation 3 : On the horizon the non-zero values of $\delta R^a_{~bcd}$ are
\be
\delta R^{V}_{~UUV}&=&-\delta R^V_{~UVU}=-{2\alpha(x)A'\over A}\,\delta (U)\\
\delta R^{V}_{~iUj}&=&-\delta R^{V}_{~ijU}=\delta(U) \bar\nabla_i\bar\nabla_j\alpha(x) -{1\over2A}\bar g_{ij}h'\alpha(x)\delta(U)\\
\delta R^{i}_{~UjU}&=&-\delta R^{i}_{~UUj}={h'\alpha(x)\over 2h}\,\delta (U)\,\delta^i_j+{A\delta (U)\over h}\,\bar\nabla^i\bar\nabla_j\alpha(x)\\
\delta R^{i}_{~UjU}&=&-\delta R^{i}_{~UUj}={h'\alpha(x)\over 2h}\,\delta (U)\,\delta^i_j+{A\delta (U)\over h}\,\bar\nabla^i\bar\nabla_j\alpha(x)
\ee

Relation 4 : On the horizon the non-zero values of $\delta R_{ab}$ are $\delta R_{UU}$ and $\delta R=0$.
\be
\delta R_{UU}&=&\Big( {2A'\over A}+\sum_S {d^{(S)}}{h'^{(S)}\over 2 h^{(S)}}\Big)\,\alpha(x)\delta(U)+\delta(U)\sum_S {A\over h^{(S)}}~\bar\Delta^{(S)}\alpha(x)\\
\delta R&=&0
\ee
where the Laplacian is defined by
\be
\bar\Delta^{(S)}~\alpha(x)={1\over \sqrt {\bar g^{(S)}}} ~\partial_i^{(S)} \Big(\sqrt {\bar  g^{(S)}} ~ {\bar g^{(S)ij}}~ \partial^{(S)}_j  \alpha(x)\Big)
\ee
and $\bar \nabla_i$ is the covariant   derivative in the space with metric $\bar g_{ij}^{(S)}$. With the helps of relation 1 $\sim$ relation 4, we can  prove following relation.  \\

Relation 5 :   On the horizon 
\be
\delta (R_{abcd}R^{abcd})=\delta (R_{ab}R^{ab})=\delta (R^2)=\delta (\Box R)\,\dot=\,0
\ee 
Last relation  plays a central role to obtain the simplified shock wave equation of quadratic gravity in below.
\subsection{Simplified Shock Wave Equation}
Using these  relations  we  begin to evaluate   ${\sf G}^{(1)}_{UU}+ 2\,{\sf G}^{(0)}_{UV} \, \alpha(x) \,\delta(U)$ in qradratic gravity theory.  After explicitly expansion we find that 
\be
&&{\sf G}^{(1)}_{UU}+2\,{\sf G}^{(0)}_{UV}\, \alpha(x) \,\delta(U)\nn\\
\label{totalL1}
&=&2\alpha \delta (R_{Ucde}R_U^{~cde}) +2(2\alpha+\beta)\delta (R_{UcUd}R^{cd})-4\alpha \delta (R_{Uc}R_U^{c})\nn\\
&&+2\gamma \delta (R_{UU}R)+(4\alpha+\beta)\delta (\Box R_{UU})-(2\alpha+\beta+2\gamma)\delta (\nabla_U\nabla_U R)\nn\\
&&-{1\over 2}\delta  \Big[g_{UU} \Big(( \alpha R_{abcd}R^{abcd}+\beta R_{ab}R^{ab} +\gamma R^2) -(\beta+4\gamma)\Box R \Big)\Big]\nn\\
&&+2\Big\{2\alpha  R_{Ucde}R_V^{~cde} +2(2\alpha+\beta) R_{UcVd}R^{cd}-4\alpha  R_{Uc}R_V^{c}\nn\\
&&+2\gamma  (R_{UV}R)+(4\alpha+\beta) (\Box R_{UV})-(2\alpha+\beta+2\gamma) (\nabla_U\nabla_V R)\nn\\
&&-{1\over 2} \Big[g_{UV} \Big(( \alpha R_{abcd}R^{abcd}+\beta R_{ab}R^{ab} +\gamma R^2) -(\beta+4\gamma)\Box R \Big)\Big]\Big\} \alpha(x) \delta(U)\\
\label{totalL2}
&=&2\alpha \delta (R_{Ucde}R_U^{~cde}) +2(2\alpha+\beta)\delta (R_{UcUd}R^{cd})-4\alpha \delta (R_{Uc}R_U^{c})\nn\\
&&+2\gamma \delta (R_{UU}R)+(4\alpha+\beta)\delta (\Box R_{UU}) -(2\alpha+\beta+2\gamma) \delta (\nabla_U\nabla_U R)\nn\\
&&+2\Big\{2\alpha  R_{Ucde}R_V^{~cde} +2(2\alpha+\beta) R_{UcVd}R^{cd}-4\alpha  R_{Uc}R_V^{c}\nn\\
&&+2\gamma  (R_{UV}R)+(4\alpha+\beta) (\Box R_{UV})-(2\alpha+\beta+2\gamma) (\nabla_U\nabla_V R)\Big\} \alpha(x)\delta(U)
\ee
To obtain the last relation we have used the properties of Proposition 5 to conclude that  the operator $\delta$ in first bracket of \eqref{totalL1} only produces  $\delta g_{UU}$. After substituting the explicitly  forms of $\delta g_{UU}=2A(UV) \alpha(x)\delta(U)$ in first bracket and $g_{UV}=A(UV)$  in second bracket we see that they are canceled to each other and we have the last relation \eqref{totalL2}. \\

Eq.\eqref{totalL2} has six zero-order terms and six first-order terms.  It is interesting to see that we could furthermore simplify it to contain only six first-order terms in another forms.  

 Using the metric properties in \eqref{metric0} and \eqref{metric1} we can find the following  simple relation, which  can be applied to  any tensor $F_{UU}$ :
\be
\label{FUU}
\delta F_{UU}&=&\delta(g_{Ua}F^a_{~U})=(\delta g_{Ua}) F^a_{~U}+g_{Ua}\delta(F^a_{~U})=(\delta g_{UU})F^U_{~U} +g_{UV}\delta(F^V_{~U})\nn\\
&=&(\delta g_{UU})g^{UV}F_{VU}+g_{UV}\delta(F^V_{~U}) =-2F_{VU}\alpha(x)\delta(U)+g_{UV}\delta(F^V_{~U})
\ee
After identifying $F_{UU}$ as $R_{Ucde}R_U^{~cde},~R_{UcUd}R^{cd},\cdot\cdot\cdot$ and substituting them into equation \eqref{totalL2} we find that 
\be
\label{totalL3}
{\sf G}^{(1)}_{UU}+2\,{\sf G}^{(0)}_{UV}\, \alpha(x) \,\delta(U)
&=&g_{UV}\Big(2\alpha \delta (R^V_{~cde}R_U^{~cde}) +2(2\alpha+\beta)\delta (R^V_{~cUd}R^{cd})-4\alpha \delta (R^V_{~c}R_{~U}^{c})\nn\\
&&+2\gamma \delta (R^V_{~U}R)+(4\alpha+\beta)\delta (\Box R^V_{~U}) -(2\alpha+\beta+2\gamma) \delta (\nabla^V\nabla_U R)\Big)~~~~~~~
\ee
Now,  it remains only six terms. Among them four terms are quadratic curvature and two terms are derivative of  curvature.   With the help of appendix and  after calculations we  collect the formulas of these six terms  in below.
\subsection{Six Formulas}
Using the proposition 1 $\sim$ proposition 4 we can find following four  formulas :

 Formula 1 : 
\be
\label{4-1}
\delta(R^V_{~bcd}R_U^{~bcd})&\dot=&\sum_S{d^{(S)}\alpha(x)(h'^{(S)})^2\over A^2\, (h^{(S)})^2}\delta(U) -2\delta(U){h'^{(S)}\over A\,(h^{(S)})^2}\bar\Delta^{(S)} \alpha(x)
\ee

Formula 2 : 
\be
\label{4-2}
\delta (R^V_{~cUd}R^{cd})&\dot=&\sum_{S}{\delta(U)\over  (h^{(S)})^2}\Big(\bar R^{(S)}_{ij} \bar\nabla^{(S)i}\bar\nabla^{(S)j}\alpha(x)  -{1\over2A}\bar R^{(S)}h'^{(S)} \alpha(x)\Big)\nn\\
&&+{\delta(U)\over Ah^{(S)}}\Big({h'^{(S)}\over h^{(S)}}-{A'\over A} \Big)\Big({d^{(S)}h'^{(S)}\over2A}\alpha(x)- \bar\Delta^{(S)}\alpha(x)\Big)~~
\ee

Formula 3 : 
\be
\label{4-3}
\delta (R^V_{~c}R_U^{c})
&\dot=& {\delta(U)\over A^2}\Big(\sum_S {{d^{(S)}}h'^{(S)}\over h^{(S)}}\,\alpha(x)-\sum_S {2A\over h^{(S)}}~\bar\Delta^{(S)}\alpha(x)\Big)\nn\\
&&\times\Big({A'\over A}+\sum_{\tilde S}{{d^{(\tilde S)}}\,h'^{(\tilde S)}\over 2h^{(\tilde S)}}\Big)
\ee

Formula 4 : 
\be
\label{4-4}
\delta (R^V_{~U}R)
&\,\dot=\,&{\delta(U)\over A}\Big(\sum_S }{{d^{(S)}h'^{(S)}\over  h^{(S)}}\,\alpha(x)-\sum_S {2A\over h^{(S)}}~\bar\Delta^{(S)}\alpha(x)) \Big)\nn\\
&&\times \Big({A'\over A^2}-\sum_{\tilde S} {\bar R^{(\tilde S)}\over 2h^{(\tilde S)}}+{d^{(\tilde S)}\,h'^{(\tilde S)}\over A\,h^{(\tilde S)}}\Big)
\ee
We perform more calculations in appendix to obtain  following two formulas :

Formula 5 : 
\be
\label{6-1}
\delta (\Box R^V_{~U})
&\dot=&\Big(\sum_{S,\tilde S} {\bar\Delta^{(S)}\bar\Delta^{(\tilde S)}\alpha(x)\over h^{(S)}h^{(\tilde S)}}\Big)\,\delta (U)-{1\over A}\Big({2A'\over A}+\sum_S{d^{(S)}h'^{(S)} \over h^{(S)}}\Big)\Big(\sum_S {\bar\Delta^{(S)} \alpha(x) \over h^{(S)}}\Big)\,\delta (U)\nn\\
&&+{1\over A^2}\left[\Big(\sum_S {d^{(S)}h'^{(S)} \over 2 h^{(S)}}\Big)^2+\sum_S d^{(S)}\Big({3A'h'^{(S)}\over Ah}-{2h''^{(S)} \over h^{(S)}}+{(h'^{(S)})^2\over (h^{(S)})^2} \Big)\right] \alpha(x)\,\delta (U)\nn\\
\ee

Formula 6 : 
\be
\label{6-2}
&&\delta (\nabla^V\nabla_U R)\,\dot=\,-\Big[\sum_S{1\over (h^{(S)})^2}\,(\bar\partial^i\alpha(x))(\bar\partial_i\bar R^{(S)})\Big]\,\delta(U)\nn\\
&&+{1\over A}\Big[{6(A')^2\over A^3}-{4A''\over A^2}-\sum_S\Big({\bar R^{(S)}h'^{(S)}\over (h^{(S)})^2 }+{4d^{(S)}h''^{(S)} \over Ah^{(S)}}-{2d^{(S)}A'h'^{(S)} \over A^2h^{(S)}} +{d^{(S)}(d^{(S)}-7)(h'^{(S)})^2 \over 2A(h^{(S)})^2}\Big)\Big]\alpha(x)\delta(U)\nn\\
\ee 

\subsection{Butterfly Velocity in Anisotropic Space of Gauss-Bonnet  Gravity}
We now apply above formulas to the simplest case of Gauss-Bonnet gravity in which  $\alpha=\gamma_{GB},~\beta=-4\gamma_{GB},~\gamma=\gamma_{GB}$.   The equation \eqref{totalL3} now has a simple form
\be
\label{mainresult2}
&&{\sf G}^{(1)}_{UU}+2\,{\sf G}^{(0)}_{UV}\, \alpha(x) \,\delta(U)\nn\\
&=&2g_{UV}\,\gamma_{GB}\,\Big( \delta (R^V_{~cde}R_U^{~cde}) -2\delta (R^V_{~cUd}R^{cd})-2 \delta (R^V_{~c}R_{~U}^{c})+\delta (R^V_{~U}R)\Big)\nn\\
&=&2\delta(U)\,\gamma_{GB}\,\Big[-2\Big(\sum_{S}{A\over  (h^{(S)})^2}\bar R^{(S)}_{ij} \bar\nabla^{(S)i}\bar\nabla^{(S)j}\alpha(x)-{\bar R^{(S)}h'^{(S)} \over 2  (h^{(S)})^2}\alpha(x)\Big)\nn \\
&&+\Big(\sum_S {A\over h^{(S)}}~\bar\Delta^{(S)}\alpha(x)-{d^{(S)}h'^{(S)}\over  2h^{(S)}}\,\alpha(x)\Big) \Big(\sum_{\tilde S} {\bar R^{(\tilde S)}\over h^{(\tilde S)}}\Big)\Big]
\ee
Let us make three comments about the result : 

1. In the case of isotropic space we can remove the summations over $S$ and $\tilde S$, then formula \eqref{mainresult2} reproduces the formula  (4.18)  in our previous paper \cite{Huang1710}. 

2. Dues to the appearance of double summations   the formula \eqref{mainresult2} is not just that  adding a simple summation over $S$ to the formula (4.18)  in \cite{Huang1710}, in which only isotropic space was analyzed .  Thus the shock wave equation in anisotropic space is a non-trivial extension of that in  isotropic space.

3.  Consider the  the planar, spherical, or hyperbolic black hole metric in \eqref{BH1} and using  the relations \eqref{BH2} and \eqref{BH3} we find that 
\be
{A\over  (h^{(S)})^2}\bar R^{(S)}_{ij} \bar\nabla^{(S)i}\bar\nabla^{(S)j}\alpha(x)-{\bar R^{(S)}h'^{(S)} \over 2  (h^{(S)})^2}\alpha(x)=k^{(S)}(d^{(S)}-1) \Big({A\over h^{(S)}}~\bar\Delta^{(S)}\alpha(x)-{d^{(S)}h'^{(S)}\over  2h^{(S)}}\,\alpha(x)\Big)~~
\ee
 Substituting this relation into \eqref{mainresult2} we see that the shock wave equation of Einstein-Gauss-Bonnet gravity and that of Einstein gravity, i.e. \eqref{Einstein}, obey the same differential equation when the space is isotropic.  The double summations  in the formula \eqref{mainresult2} will ruin this property when the space is anisotropic.  
 
   Thus, we conclude that {\it in the D-dimensional planar, spherical or hyperbolic black hole spacetime  the Einstein-Gauss-Bonnet gravity has the same shock wave equation as that in Einstein gravity if  and only if the space is isotropic}.
\section{Butterfly Velocity in  Isotropic Spaces of Quadratic Gravity}
\subsection{Formula}
In this section  we will  find the simplified formula of shock wave equation for the quadratic gravity in the following spacetime\footnote {Through coordinate transformation the space can become a  general form : $ds^2=-a(\tilde r)\tilde f(\tilde r)dt^2 +{d\tilde r^2\over b(\tilde r)\tilde f(\tilde r)} + \tilde h(\tilde r)\bar g_{ij}(x)dx^idx^j$  and property found in this paper is very general}
\be
\label{metricS}
ds^2&=&-N_\sharp^2f(r)dt^2+{dr^2\over f(r)}+ h(r)\sum_{i,j=1}^d\bar g_{ij}(x)dx^idx^j
\ee
The constant $N_\sharp^2$ is  introduced to make the metric to be AdS space  asymptotically.  We  consider 2+d dimensional planar, spherical or hyperbolic black holes.  The general metric is
\be
\label{BH1}
\bar g_{ij}(x)dx^idx^j&=&\left\{
 \ba {cc}
d\theta_1^2+d\theta_2^2+ \cdot\cdot\cdot+d\theta_d^2,&k=0\nn\\
 d\theta_1^2+\sin^2\theta_1(d\theta_2^2+\sin^2\theta_2(d\theta_3^2+ \cdot\cdot\cdot+\sin^2\theta_{d-1}d\theta_d^2),&k=1\\
 d\theta_1^2+\sinh^2\theta_1(d\theta_2^2+\sin^2\theta_2(d\theta_3^2+ \cdot\cdot\cdot+\sin^2\theta_{d-1}d\theta_d^2),&~~k=-1\nn\\
 \ea
 \right.
 \\
\ee
which implies
\be
\label{BH2}
\bar R^{ij} \bar\nabla_i\bar\nabla_j\alpha(x)&=&k(d-1)\bar\Delta\alpha(x)\\
\label{BH3}
\bar R&=&kd(d-1)
\ee
Note that the shock wave equation has two parts, one is from Einstein gravity (EG) and another is from quadratic gravity (QG).  The results are    
\be
{\sf G}^{(1)}_{UU}+2{\sf G}^{(0)}_{UV} \alpha(x)\delta(U)&=&\Big[{\sf G}^{(1)}_{UU}+2{\sf G}^{(0)}_{UV}~ \alpha(x) ~\delta(U)\Big]_{EG}+\Big[{\sf G}^{(1)}_{UU}+2{\sf G}^{(0)}_{UV}~ \alpha(x) ~\delta(U)\Big]_{QG}~~~
\ee
where
\be
\label{Deltaalpha}
\Big[{\sf G}^{(1)}_{UU}+2{\sf G}^{(0)}_{UV}~ \alpha(x) ~\delta(U)\Big]_{EG}
&=&{A\over  r_H^2}\Big(\Delta\alpha(x)-{d\over 2} \,r_H\, f'(r_H)\alpha(x)\Big)\,\delta(U)\\
\nn\\
\Big[{\sf G}^{(1)}_{UU}+2{\sf G}^{(0)}_{UV}~ \alpha(x) ~\delta(U)\Big]_{QG}&=& A\left(2\alpha \delta (R^V_{~cde}R_U^{~cde}) +2(2\alpha+\beta)\delta (R^V_{~cUd}R^{cd})-4\alpha \delta (R^V_{~c}R_{~U}^{c})\right.\nn\\
&&+2\gamma \delta (R^V_{~U}R)+(4\alpha+\beta)\delta (\Box R^V_{~U}) -(2\alpha+\beta+2\gamma) \delta (\nabla^V\nabla_U R)\Big)~~~~~~~~
\ee
 and 
\be
\delta(R^V_{~bcd}R_U^{~bcd})
&=&-{2f'(r_H)\over r_H^3}\Big(\bar\Delta\alpha(x)-{d\over 2}r_H\,f'(r_H)\,\alpha(x) \Big)\\
\delta (R^V_{~cUd}R^{cd})
&=&{r_H^2f''(r_H)-2r_Hf'(r_H)-2(1-d)dk\over 2r_H^4}\Big(\bar\Delta\alpha(x)-{d\over 2}r_H\,f'(r_H)\,\alpha(x) \Big)\\
\delta (R^V_{~c}R_U^{c})
&=&-{r_Hf''(r_H)+df'(r_H)\over r_H^3}\Big(\bar\Delta\alpha(x)-{d\over 2}r_H\,f'(r_H)\,\alpha(x) \Big)\\
\label{GBP}
\delta (R^V_{~U}R) 
&=&-{r_H^2f''(r_H)+2dr_Hf'(r_H)+(1-d)dk\over r_H^4}\Big(\bar\Delta\alpha(x)-{d\over 2}r_H\,f'(r_H)\,\alpha(x) \Big)~~~~~~~
\ee
which are  calculated from \eqref{4-1}, \eqref{4-2}, \eqref{4-3} and \eqref{4-4} respectively.  And
\be
\delta (\Box R^V_{~U})
&=&{\bar\Delta\bar\Delta\alpha(x)\over r_H^4}-{f'(r_H)\over r_H^3}\Big(d+r_Hf''(r_H)\Big) \bar\Delta\alpha(x) +{df'(r_H)\over 4r_H^2}\Big(2r_Hf''(r_H) +df'(r_H)\Big)\alpha(x)\nn\\
\\
\nn\\
\delta (\nabla^V\nabla_U R)\,
&=&-{\alpha(x)\,f'(r_H)\over 2r_H^3}\Big(2(d-1)d\,k +d\,r_H((d-3) f'(r_H) +2r_Hf''(r_H)) +r_H^3f'''(r_H)\Big)~~~~~~~~~~~
\ee 
which are  calculated from \eqref{6-1} and \eqref{6-2} respectively.   Note that  \eqref{4-1}, \eqref{4-2}, \eqref{4-3} and \eqref{4-4} have a common  factor $``\bar\Delta\alpha(x)-{d\over 2}r_H\,f'(r_H)\,\alpha(x) ``$.  This factor is just that  appears in the  Einstein gravity \eqref{Einstein}. 

   We now apply these formula to the  spacetime \eqref{metricS} with \footnote{ This is the planar black solution.}
\be
\label{metricF}
f(r)&=&r^2\left(1-\Big({r_0\over r}\Big)^{d+1}+\delta+\eta\Big({r_0\over r}\Big)^{2(d+1)}\right),~~~~h(r)=r^2\\
\delta&=&{(d - 2)\over d} \Big[(d + 1) \Big((d + 2) \gamma + \beta\Big) + 2 \alpha\Big],~~~~\eta=(d - 1) (d - 2) \alpha\\
N_\sharp^2&=&1+\delta
\ee
The black hole horizon and temperature \footnote{The  temperature form in \eqref{T} is shown in \cite{Kats0712}, which is that without factor $N_\sharp$.} are
\be
r_H&=&r_0\,\Big(1-{\delta+\eta\over d+1}\Big)\\
\label{T}
T&=&{(d+1)r_H\over 4\pi}\left[1-\gamma(d-2)(d-1) +{(d-2)\Big((d+1)(\alpha (d+2)+\beta) +2\gamma\Big)\over 2d}\right]
\ee
Above metric was first derived in \cite{Kats0712}.  It had been used by \cite{Kats0712, Brigante0712} to show the viscosity bound violation and shear sum rule in higher derivative gravity theories \cite{ Chowdhury1711}.  We will use this metric to study the effect of quadratic gravity on the butterfly velocity. 

  After the calculation the shock wave equation  \eqref{sweq} becomes
\be
C_2\,\bar\Delta\bar\Delta\alpha(x)+C_1\,\bar\Delta\alpha(x)+C_0\,\alpha(x)&=&E\,e^{2\pi t/\beta}~a(x)
\ee
where
\be
C_2&=&(4 \alpha + \beta){1\over r_H^4}\\
C_1&=& -(1 + d)^2 (4 \alpha + \beta)+{1\over r_H^2}\Big( 1 - 2 (-2 + d + 3 d^2) \alpha - 4 \gamma - 2 d^2 (\beta + \gamma) + 2 d (\beta + 3 \gamma)\Big)\\
C_0&=&(1 + d) \Big[ \Big(-2 d +2 (1 + d) (4 + d^2)\Big) \alpha + (1 + d) (4 + d (2 + d)) \beta + 2 (1 + d) (2 + d)^2 \gamma\Big]~~~~~~~~
\ee
The appearing of the term $\bar\Delta\bar\Delta\alpha(x)$, which is fourth-order derivative of $\alpha(x)$,  is the general property after introducing the quadratic gravity. 

To proceed we can follow the paper  \cite{Alishahiha1610} to find the  two butterfly velocities  therein. In general the solution can be written as
\be
\alpha(x)\sim e^{{2\pi\over \beta}\big(t-t_*-{|x|\over v_B^{(1)}}\big)}-{v_B^{(2)}\over v_B^{(1)}}\,e^{{2\pi\over \beta}\big(t-t_*-{|x|\over v_B^{(2)}}\big)}
\ee
where the butterfly velocity is defined by  \cite{Shenker1306}
\be
v_B^{(i)}=N_\sharp\,{2\pi T\over M^{(i)}}
\ee
$M^{(i)}$ are calculated from the following equation
 \be
C_2\,\bar\Delta\bar\Delta\alpha(x)+C_1\,\bar\Delta\alpha(x)+C_0\,\alpha(x)&=&C_2\,\Big(\bar\Delta-(M^{(1)})^2\Big)\Big(\bar\Delta-(M^{(2)})^2\Big)
\ee
The details are described by Alishahiha et. al. in  \cite{Alishahiha1610}.

After calculation the final formula of the holographic butterfly velocity propagating in the space  \eqref{metricF} becomes
\be
\label{finalVB1}
v_B^{(1)}&=&\sqrt{d+1\over 2 d}\Big[1- 8\pi^2 (\beta+4\alpha) \,T^2  -{1\over 2} (d-2) \Big((d-1) \alpha + ( d+1)(\beta+4\alpha)+ (3d+1)(\gamma-\alpha) \Big)\Big]\nn\\
&&~~~~~~~~~~~~+{\cal O}\Big((\alpha,\beta,\gamma)^2\Big)\\
\label{finalVB2}
v_B^{(2)}&=&{(d+1)\sqrt{-(\beta+4\alpha)}\over 2 }+{\cal O}\Big((\alpha,\beta,\gamma)^{3/2}\Big)
\ee
in which $(\alpha,\beta,\gamma)^2$ represents any function of second order of variables $\alpha,\beta,\gamma$.   We now use above relations to discuss the various properties of butterfly velocity in quadratic gravity.
\subsection{Example : Butterfly Velocity in Quadratic Gravity}
1. The second velocity becomes zero if $4\alpha+\beta=0$. This is the result that $C_2=0$ when $4\alpha+\beta=0$ and the shock wave equation becomes second-order derivative differential equation. It is interesting to see that {\it second velocity  can appear only if  $4\alpha+\beta<0$}. 

2. In the case of   $\beta+4\alpha=0$, this including the $R+\gamma R^2$  gravity,  the second  velocity is zero and the first velocity becomes
\be 
v^{(1)}_B&=&\sqrt{d+1\over 2 d}\Big[1-{1\over 2} (d-2) \Big((d-1) \alpha + (3d+1)(\gamma-\alpha) \Big)\Big]\nn\\
\ee
In this case the butterfly velocity in D=4  black hole, i.e.  d=2,  does not have   correction by the quadratic curvatures.

3.  The quantities $(\beta+4\alpha)$ and $(\gamma-\alpha)$ in eq.\eqref{finalVB1}  measure the deviations from the Gauss-Bonnet gravity.  The case of  both quantities being vanish corresponds to the Einstein-Gauss-Bonnet gravity and
\be 
\label{ratio}
v^{(GB)}_B(\alpha)=\left[1-{1\over 2}\alpha (d-1)(d-2)  \right]\,v^{(GB)}_B(0),~~~~~~~v^{(GB)}_B(0)=\sqrt{d+1\over 2 d}
\ee 
which was first found in  \cite {Roberts1409}. The factor $\left[1- {1\over 2} \alpha (d-1)(d-2) \right]$ is from the constant value of $N_\sharp$, which defined in \eqref{metricS},  while the another factor  $\sqrt{d+1\over 2 d}$ is from the shock wave equation in the Einstein gravity.  

4.  When d=2, i.e. D=4,  the the butterfly velocities are functions of $(\beta+4\alpha)$ which is zero in Gauss-Bonnet gravity. This reveals that D=4  Gauss-Bonnet term is topological. 

5. In the case of   $R+\gamma R^2$ gravity  the second  velocity is zero and the first velocity becomes
\be 
v^{(R^2)}_B&=&\sqrt{d+1\over 2 d}\Big[1-{(d-2)(3d+1)\gamma\over2}\Big]
\ee
This means that for the D=4 planar black hole the $ R^2$ gravity  does not give any correction to the butterfly velocity.  Otherwise  the correction maybe positive or negative, depending on the values  of $\gamma$.

6. In the  Einstein-Conformal  gravity in which  $\beta=-2\alpha$ and $\gamma={1\over 3}\alpha$ the second butterfly velocity becomes
\be 
v^{(1)}_B&=&\sqrt{d+1\over 2 d}\Big[1-\Big(16\pi^2 T^2+{(d-2)(3d+1)\over6}\Big)\,\alpha\Big]
\ee
which shows a different behavior form that in $R^2$ gravity, since that for the D=4 planar black hole the conformal  gravity  will correct the butterfly velocity.

7. At high temperature  we have a simple relation
\be
\label{HTV}
v_B^{(1)}&\approx&\sqrt{d+1\over 2 d}\left[1-8\pi^2(\beta+4\alpha) T^2+{\cal{O}}(T^{(0)})\right]
\ee
which is independent of the values of  $\gamma$.  This means that, at high temperature the correction of butterfly velocity is independent of  $R^2$ term.  $R^2$ term  can correct the  butterfly velocity at order ${\cal O}(T^0)$.

8. At low temperature we have a simple relation
\be
\label{LTV}
v_B^{(1)}&\approx&\sqrt{d+1\over 2 d}\left[1-{1\over 2} (d-2) \Big((d-1) \alpha + ( d+1)(\beta+4\alpha)+ (3d+1)(\gamma-\alpha) \Big)\right]
\ee
Comparing above equation to the case of high temperature expansion we see that,  depending on the values of $\alpha$, $\beta$, $\gamma$ and $d$, the velocity correction from the quadratic gravity may be from positive to negative or from negative to positive while increasing the temperature. \\

 Finally, We  can directly apply the general  formula derived in previous subsection to a simple case in which the metric solution is the  Schwarzschild-AdS black hole solution   
\be
ds^2=-r^2\left(1-{r_H^3\over r^3}\right)dt^2+{dr^2\over r^2\left(1-{r_H^3\over r^3}\right)}+r^2(dx^2+dy^2)
\ee
since that any solution of the pure Einstein theory continues to be a solution of the theory with
the quadratic modifications.. This  is the spacetime  considered by Alishahiha et. al. in \cite{Alishahiha1610}. The fourth order differential equation of shock wave equation becomes
\be
&&{4\alpha+\beta\over r_H^4}\,\bar\Delta\bar\Delta\alpha(x)+{1-3(8\alpha+4\beta+3r_H^2(4\alpha+\beta)+8\gamma)\over r_H^2}\,\bar\Delta\alpha(x)\nn\\
&&~~~~~~~~~~~+(-3+36\alpha+27\beta+72\gamma)\,\alpha(x)=E\,e^{2\pi t/\beta}~a(x)
\ee
After the calculations the butterfly velocities becomes 
\be
v_B^{(1)}&=&{\sqrt3\over 2}\left(1-{8\pi^2(4\alpha+\beta)\over |1-6\beta-24\gamma|}\,T^2+{\cal O}(T^4)\right)\\
v_B^{(2)}&=&{3\over 2} \sqrt{(4\alpha+\beta)\over 12\alpha+9\beta+24\gamma-1}\,\left(1+{8\pi^2(4\alpha+\beta)\over |1-6\beta-24\gamma|}\,T^2+{\cal O}(T^4)\right)
\ee
which consists with  eq.(22) in \cite{Alishahiha1610} when $\alpha=T=0$.  The second velocity is real if $ {(4\alpha+\beta)\over 12\alpha+9\beta+24\gamma-1}>0$. The condition reduces to $4\alpha+\beta<0$ for small values of $\alpha,\beta,\gamma$.
\subsection{Example : Butterfly Velocity in  Gauss-Bonnet Massive  Gravity}
 Since that ${\cal U}_i$ in \eqref{MG} are functions of metric $g_{\mu\nu}$ and reference metric  $f_{\mu\nu}$ while do not depend on the Riemann curvature we can regard them  as some kinds of the extra matter fields. Thus the formula derived in previous section, which analyze the variation with respect to the Riemann curvature,  can be directly applied.   Thus we conclude that :
 
 {\it In the D-dimensional planar, spherical or hyperbolic black hole spacetime  the Einstein-Gauss-Bonnet massive gravity has the same shock wave equation as that in Einstein gravity if  and only if the space is isotropic.}
 
Therefore, in the case of  isotropic space we can quickly  calculate the butterfly velocity in  the Gauss-Bonnet massive  gravity theories, following the method  described in our previous paper \cite{Huang1710},  i.e. \eqref{mainvelocity}. We consider  the  (d+2) dimensional Maxwell-Gauss-Bonnet massive gravity. The Lagrangian is described by \eqref{MG}   where  ${\cal L}_{\rm matters}$ is the Maxwell field and we add the Gauss-Bonnet curvature with coefficient $\alpha$.  The charged black hole solution found in \cite{Hendi1507} is
 \be
ds^2&=& -N_\sharp^2f(r)dt^2+{dr^2\over f(r)}+r^2 \delta_{ij}dx^idx^j,~~~~~~~~i,j=1,2,3....d\\
F_{tr}&=&{Q\over r^{d}}\\
f(r)&=&k+{r^2\over 2\alpha\,d_3d_4}\,\left\{1-\sqrt{1+{ 8\alpha\,d_3d_4\over d_1d_2}\left[\Lambda+{ d_1d_2m_0\over 2r^{d_1}}-{Q^2\,d_1\over d_3r^{2d_2}}+\Upsilon\right]}\right\}\\
\Upsilon&=&-m^2\,d_1d_2\left[{d_3 d_4c^4c_4\over 2r^4}+{ d_3c^3c_3\over 2r^3}+{ c^2c_2\over 2r^2}+{cc_1\over 2d_2r}  \right]
\ee
The reference metric  is chosen to be $f_{\mu\nu}=(0,0,c^2\delta_{ij})$.  The notation $d_i=d+2-i$ is used. The constant  $N_\sharp^2$  is
\be
N_\sharp^2 &=&{1\over2}\,\left(1+\sqrt{1-2\alpha(d-1)(d-2)}\right)
\ee
after substituting the conventional value of $\Lambda=-{d(d+1)\over 4}$.  The horizon defined by  $f(r_H)=0$ leads to relation
\be
1+{2k\alpha\,d_3d_4\over r_H^2}&=&\sqrt{1+{ 8\alpha\,d_3d_4\over d_1d_2}\left[\Lambda+{ d_1d_2m_0\over 2r_H^{d_1}}-{Q^2\,d_1\over d_3r_H^{2d_2}}+\Upsilon_H\right]}
\ee
and black hole temperature is 
\be
4\pi T&=&-{2k\over r_H }+{ d_1m_0\over r_H^{d_1-1}}-{4Q^2\over d_3r_H^{2d_2-1}}-m^2\left[{4d_3 d_4c^4c_4\over r_H^3}+{ 3d_3c^3c_3\over r_H^2}+{2 c^2c_2\over r_H}+{cc_1\over d_2}  \right]
\ee
Above relation can be used to express $r_H$ as a function of temperature while  does not explicitly depend on  $\alpha$.   Therefore, after using  the basic formula $v_B=N_\sharp\,\sqrt{4\pi T\over 2 d r_H}$  the butterfly velocity has an exact relation
\be
v_B^m(\alpha)&=&\left[{1\over2}\,\left(1+\sqrt{1-2\alpha(d-1)(d-2)}\right)\right]^{1/2}\,v_B^m(0)
\ee
The ratio between $v_B^m(\alpha)$ and $v_B^m(0)$ had appeared in previous literature \cite{Roberts1412, Huang1710} and \eqref{ratio}\footnote{ \eqref{ratio} is leading order of $\alpha$}.  Since this paper is to see how the quadratic curvature affect the butterfly velocity the property of  $v_B^m(0)$ is left to reader to analyze. Note that above relation can also appear in 
Gauss-Bonnet massive gravity with Born-Infeld electrodynamics \cite{Hendi1510} or in the presence of power-Maxwell field \cite{Hendi1708}, since that  $r_H$ in these cases still are  a function of temperature while  does not explicitly depend on  $\alpha$. 

\section{Conclusions}
In this paper we continue  previous work \cite{Huang1710} to study the butterfly velocity in general quadratic  gravity with Lagrangian ${\cal L}= \alpha R_{\mu\nu\sigma\rho} R^{\mu\nu\sigma\rho}+\beta R_{\mu\nu}R^{\mu\nu}+\gamma R^2+{\cal L}_{\rm matter}$. Contrast to the case of Gauss-Bonnet theory, in which $\alpha=\gamma=-{\beta\over 4}$, the quadratic gravity can correct the  shock wave equation.   After the detailed tensor calculations the general formula of shock wave equation in the  general  anisotropic spacetime is derived.  We use the formula to prove that in the D-dimensional planar, spherical or hyperbolic black hole spacetime  the  shock wave equation in the Einstein-Gauss-Bonnet gravity has the same form  as that in Einstein gravity only if the space is isotropic. 

    We consider the example of a simple  spacetime, which is the solution  in leading order of $\alpha$, $\beta$ and $\gamma $. We  obtain   a simple formula  of butterfly velocity in eq.\eqref{finalVB1} and eq.\eqref{finalVB2}.  Using the formula we find that the fourth-derivative shock wave equation therein could  lead to  two butterfly velocities if and only if $ 4\alpha+\beta<0 $.  We also see that the  D=4  planar black hole  does not give  correction to the butterfly velocity in the quadratic gravity with $ \beta +4\alpha=0$, which includes the $ R^2$ gravity.    We also see that, depending on the values of $\alpha$, $\beta$, $\gamma$, and the  black-hole shape the velocity correction from the quadratic gravity may be from positive to negative or from negative to positive while increasing the temperature.    The  butterfly velocity in the theory of Gauss-Bonnet massive gravity is also studied.
      
    Since our formula collected in section 4 is very general it can be applied to general anisotropic space with arbitrary matter fields.  While the application in section 5 is in a simple isotropic space it is interesting to applied it to more complex space with matters and to see how the butterfly velocity will be in there.  It  hopes that our formula is helpful in studying the butterfly velocity in quadratic gravity.
    
    We make four comments to conclude this paper. 
    
    1.    As mentioned  in section I that the value of $\alpha$ is related to the ${1\over 2N}$.  Also the values of  $\beta$ and  $\gamma$ are related the R charges of the theory.  Thus the calculations in section  5  tell us that  correction of butterfly velocity, which describes how the perturbation spreads,  may be positive or negative  depending on the values of ${1\over 2N}$, R-charges and temperature of the theory. 
    
  2.  It is known that there is a simple relation between  diffusion constant and butterfly velocity   :  $D_c\sim v_B^2T$ \cite{Blake1603}.  Thus, using this relation  we can read the property of how the diffusion constant  will depends on the values of ${1\over 2N}$, R-charges and temperature of the theory.

    3. It remains to find a simple explanation of   why the shock wave equation of Einstein gravity  does not be modified in  the Gauss-Bonnet gravity with any matters for the planar, spherical or hyperbolic black hole spacetime  in  the case of isotropic space ? 
    
    4. It is known that the bound (KSS)  violation of  viscosity  in higher derivative gravity have been  explained to relate  to the Weyl anomaly and central charge \cite{Kats0712,Brigante0712, Myers0812}. So, does butterfly velocity corrected by higher derivative gravity has any simple explanation? The investigations in \cite{Alishahiha1610} and \cite{Qaemmaqami1707} had related it to the conformal dimension and central charge. The more general property is worthy to be studied in detain.
    
    The answers to the problems could help us to understand the intrinsic properties of the quantum chaos.
\\
 \\
\appendix
\section{Exact Forms of the Christoffel symbol $\Gamma^a_{bc}$ and  $\delta \Gamma^a_{bc}$}
Tensor Calculations of this paper begin with the  exact forms of the Christoffel symbol $\Gamma^a_{bc}$ and $\delta\Gamma^a_{bc}$ in the metric \eqref{metric0}, which we present in below.

{\bf Properties of  $\Gamma^a_{bc}$} : Non-trivial values of the Christoffel symbols  $\Gamma^a_{bc}$ are
\be
\Gamma^U_{UU}&=:&{VA'(UV)\over A(UV)}\dot=0, ~~~~\Gamma^U_{ij}=:-{Uh'(UV)\over 2A(UV)}\,\bar g_{ij} \dot=0\\
\Gamma^V_{VV}&=:&{UA'(UV)\over A(UV)}\dot=0, ~~~~\Gamma^V_{ij}=:-{Vh'(UV)\over 2A(UV)}\,\bar g_{ij}\dot=0\\
\Gamma^i_{Uj}&=&\Gamma^i_{jU}=:{Vh'(UV)\over 2h(UV)}\,\delta^i_j \dot=0\\
\Gamma^i_{Vj}&=&\Gamma^i_{jV}=:{Uh'(UV)\over 2h(UV)}\,\delta^i_j \dot=0\\
\Gamma^i_{jk}&=&{1\over2}g^{im}(g_{mj,k}+g_{mk,j}-g_{jk,m})=\bar \Gamma^i_{jk} \dot \ne 0
\ee
 and all other components are exact zero.   Note the  notation $\dot=$ is used to represent the value that calculated on the horizon while the notation $=:$ is used to emphasize that  the  relation has not yet been put on horizon.  Values of not being on horizon are necessary in some calculations.

{\bf Properties of  $\delta\Gamma^a_{bc}$} : Non-trivial values of the Christoffel symbols  $\delta\Gamma^a_{bc}$ are
\be
\delta\Gamma^U_{UU}&=:&{UA'\over A}\alpha(x)\delta(U)\,\dot=0\\
\delta\Gamma^V_{UU}&=:&{1\over U}\alpha(x)\delta(U)+{VA'\over A}\alpha(x)\delta(U)\\
\delta\Gamma^V_{UV}&=&\delta\Gamma^V_{VU}=:-{UA'\over A}\alpha(x)\delta(U)\,\dot=0\\
\delta\Gamma^V_{Ui}&=&\delta\Gamma^V_{iU}=:-\delta(U)\partial_i\alpha(x)\\
\delta\Gamma^V_{ij}&=:&{Uh'\over A}\alpha(x)\delta(U)\bar g_{ij}\,\dot=0\\
\delta\Gamma^i_{UU}&=:& g^{ij}A\delta(U) \partial_j\alpha(x)
\ee
\section{Exact Forms of  $\nabla_a\nabla_b\delta g_{cd}$}
{\bf Properties of  $\nabla_a\nabla_b\delta g_{cd}$} : On the horizon the non-zero values of $\nabla_a\nabla_b\delta g_{cd}$ are
\be
\nabla_U\nabla_U\delta g_{UU}&\dot=&-2A\alpha(x)\delta''(U)\\
\nabla_U\nabla_i\delta g_{UU}&\dot=&\nabla_i\nabla_U\delta g_{UU}\dot=-2A\delta'(U) \partial_i\alpha(x)\\
\nabla_V\nabla_U\delta g_{UU}&\dot=&4A'\alpha(x)\delta(U)\\
\nabla_i\nabla_j\delta g_{UU}&\dot =&-2A\delta(U)\bar\nabla_i\bar\nabla_j\alpha(x)+\bar g_{ij}h'\alpha(x)\delta(U)\\
\nabla_j\nabla_U\delta g_{Ui}&\dot=&h'\alpha(x)\delta (U)\,\bar g_{ij}
\ee
where  $\bar \nabla_i$ is the covariant  derivative in the space with metric $\bar g_{ij}^{(S)}$. Substituting above relations into eq.\eqref{textbook} we can obtain the propositions 3 and 4.
\section{Calculation of  $\nabla^2\delta g_{UU}$}
We also need the relation 
\be
\nabla^2\delta g_{UU}&\dot=&\partial^a  \partial_a  \delta g_{UU}-2\Gamma^b_{aU}   \partial^a \delta g_{Ub}-\Gamma^{ba}_{~~a}  \partial_b \delta g_{UU} -2\Gamma^{ba}_{~~U}  \partial_a \delta g_{bU}-2\delta g_{bU}   \partial^a \Gamma^b_{aU}\nn\\
 &\dot=&2g^{UV}\partial_U  \partial_V  \delta g_{UU}+g^{ij}\partial_j  \partial_i  \delta g_{UU}-g^{ij}\Gamma^{U}_{ij}  \partial_U \delta g_{UU}-g^{jk}\Gamma^{i}_{jk}  \partial_i \delta g_{UU}-2g^{UV}\delta g_{UU}   \partial_V \Gamma^U_{UU}\nn\\
 &\dot=&\Big({4A'\over A}+\sum_S {d^{(S)}}{h'^{(S)}\over  h^{(S)}}\Big)\,\alpha(x)\delta(U)-2\delta(U)\sum_S {A\over h^{(S)}}~\bar\Delta^{(S)}\alpha(x)
 \ee
where the Laplacian is defined by
\be
\bar\Delta^{(S)}~\alpha(x)={1\over \sqrt {\bar g^{(S)}}} ~\partial_i^{(S)} \Big(\sqrt {\bar  g^{(S)}} ~ {\bar g^{(S)ij}}~ \partial^{(S)}_j  \alpha(x)\Big)
\ee
and we have used the following property :
 \be
 \partial_U\partial_V \Big[A(UV)\delta(U)\Big]= A'\delta(U)+UA'\,\delta'(U)
 =A'\delta(U)+UA'\,{-\delta(U)\over  U}=0
 \ee
It is interesting to note that despite $\delta g_{UU}$  is not a scalar field we find a simple relation 
 \be
\nabla^2\delta g_{UU}={1\over \sqrt { g}} ~\partial_a \Big(\sqrt { g} ~  g^{ab}~ \partial_b  (\delta g_{UU})\Big)
\ee
The properties of $\delta g_{ab}=\delta g_{UU}\delta^U_a\delta^U_a$ and metric form in \eqref{metric0}  are the necessary conditions to have the above relation.
\section{Exact Forms of $R_{ab}$,  $R$, $\delta R_{ab}$ and  $\delta R$}
We need the following exact forms, which are off the  horizon, to find formulas 5 and formula 6.
\be
R_{UU}&=&-V^2\sum_S{d^{(S)}h''^{(S)}\over 2h^{(S)}}+{V^2A'\over A} \sum_S {d^{(S)}h'^{(S)}\over 2h^{(S)}} +V^2\sum_S{d^{(S)}(h'^{(S)})^2\over 4(h^{(S)})^2}\\
R_{VV}&=&-U^2\sum_S{d^{(S)}h''^{(S)}\over 2h^{(S)}}+{U^2A'\over A} \sum_S {d^{(S)}h'^{(S)}\over 2h^{(S)}} +U^2\sum_S{d^{(S)}(h'^{(S)})^2\over 4(h^{(S)})^2}\\
R_{UV}
&=&-{A'\over A} -\Big(\sum_S {d^{(S)}h'^{(S)} \over 2h^{(S)}}\Big) +UV\Big({(A')^2\over A^2}-{A''\over A}-\sum_S {d^{(S)}h''^{(S)} \over 2h^{(S)}}+\sum_S{d^{(S)}(h'^{(S)})^2\over 4(h^{(S)})^2}\Big)\nn\\
\\
R_{Ui}&=&0\\
R_{ij}&=&\bar R_{ij}-{h'\over A}\,\bar g_{ij}-UV\Big({h''\over A}+{(d^{(S)}-2)(h')^2\over 2Ah}\Big)\,\bar g_{ij}^{(S)}\\
 \label{R}
R&=&-{2A'\over A^2} +\sum_S \Big({\bar R^{(S)}\over h^{(S)} }-{2d^{(S)}h'^{(S)} \over Ah^{(S)}}\Big) \nn\\
&&+UV\Big({2(A')^2\over A^3}-{2A''\over A^2}-\sum_S {2d^{(S)}h''^{(S)} \over Ah^{(S)}}-{d^{(S)}(d^{(S)}-3)(h'^{(S)})^2\over 2A(h^{(S)})^2}\Big)
\ee
in which  we use $\bar R^{(S)}_{ij}$ and  $\bar R^{(S)}$  to denote the curvature evaluated in the  metric $ds^2=\bar g^{(S)}_{ij}(x) dx^idx^j$. Note that $d^{(S)}= {\bar g^{ij(S)}} \bar g^{(S)}_{ij}$ is the dimension of space $dx_{(S)}^i$ in \eqref{metric0}. On the horizon above results reduce to proposition  2.
\be 
\delta R_{UU}&=&\Big[{2A'\over A}+\sum_Sd^{(S)}{h'^{(S)}\over2 h} +UV\Big({2A''\over A}-{2(A')^2\over A^2}+\sum_Sd^{(S)}{A'h'^{(S)}\over Ah^{(S)}}\Big)\Big]\,\alpha(x)\,\delta (U)\nn\\
&&+\delta(U)\sum_S{A\,\bar\Delta^{(S)}\alpha(x)\over h^{(S)}}\\
\delta R_{UV}&=&U^2\Big[{(A')^2\over A^2}-{A''\over A}-\sum_Sd^{(S)}{A'h'^{(S)}\over 2 Ah^{(S)}}\Big] \alpha(x)\delta(U) \\
\delta R_{ij}^{(S)}&=&-U^2\Big[{h''^{(S)}\over A}+(d^{(S)}-2){(h'^{(S)})^2\over 2 Ah^{(S)}}\Big]\,\bar g^{(S)}_{ij} \alpha(x)\delta(U) \\
 \label{dR}
\delta R&=&\delta (g^{ab}R_{ab})=(\delta g^{VV})R_{VV}+2g^{UV}\delta (R_{UV})+g^{ij}\delta (R_{ij})\nn\\
&=&U^2\Big[{2(A')^2\over A^3}-{2A''\over A^2}-\sum_S\Big({2d^{(S)}h''^{(S)}\over Ah^{(S)}}+d^{(S)}(d^{(S)}-3){(h'^{(S)})^2\over 2 A(h^{(S))^2}}\Big)\Big]\, \alpha(x)\delta(U)\nn\\
\ee
On the horizon above results reduce to proposition  4.
\section{Calculations of $\delta (\Box R^V_{~U})$ and $\delta (\nabla^V\nabla_U R)$ }
Use the  above results  we can find the  two  curvature derivative terms : 
\subsection{Calculation of  $\delta (\Box R^V_{~U})$}
We use the  following expansion 
\be
\delta (\Box R^V_{~U})&=&\delta (\Box (g^{Va}R_{aU}))=(\delta g^{VV})(\Box R_{VU})+g^{VU}\delta (\Box R_{UU})
\ee
where
\be
\Box R_{UV}&\dot=&\partial^a  \partial_a  R_{UV}-R_{bV}   \partial^a \Gamma^b_{aU} -R_{Ub}   \partial^a \Gamma^b_{aV}\nn\\
&\dot=&2g^{UV}\partial_U  \partial_V  R_{UV}-R_{UV}g^{UV}(\partial_V \Gamma^U_{UU}+ \partial_U \Gamma^V_{VV})\nn\\
&\dot=&{6(A')^2\over A^3}-{4A''\over A^2}+\sum_S d^{(S)}\Big({A'h'^{(S)}\over A^2h}-{2h''^{(S)} \over Ah^{(S)}}+{3(h'^{(S)})^2\over 2A(h^{(S)})^2} \Big)
 \ee
Next, we expand 
\be
\delta(\Box R_{UU})=(\delta\Box) R_{UU}+\Box (\delta R_{UU}) 
\ee
in which $(\delta\Box)$ is the Laplacian operator  while the  Christoffel symbol shall be replaced by $\Gamma^a_{bc}\rightarrow \Gamma^a_{bc} +\delta \Gamma^a_{bc}$ and keeps first order in $\alpha(x)$.  The first term becomes\\
\be
(\delta\Box) R_{UU}&=&\delta g^{ab}\partial_a\partial_bR_{UU}-2\delta\Gamma^b_{aU}   \partial^a R_{Ub}
-\delta\Gamma^{ba}_{~~a}  \partial_b R_{UU}-2\delta\Gamma^{ba}_{~~U}  \partial_a R_{bU}-2R_{bU}   \partial^a \delta\Gamma^b_{aU}\nn\\
&&+2 \delta\Gamma^b_{aU}    \Gamma^{ca}_{~~U}R_{bc} +2\delta\Gamma^{ca}_{~~a}  \Gamma^b_{cU}R_{bU}+2\delta\Gamma^b_{ac}    \Gamma^{ca}_{~~U}R_{bU} \nn\\
&&+2 \Gamma^b_{aU} \delta   \Gamma^{ca}_{~~U}R_{bc} +2\Gamma^{ca}_{~~a}  \delta\Gamma^b_{cU}R_{bU}+2\Gamma^b_{ac}  \delta  \Gamma^{ca}_{~~U}R_{bU}\nn\\
&\dot=&\Big({8A''\over A^2}-{8(A')^2\over A^3}+\sum_S {2d^{(S)}h''^{(S)} \over Ah^{(S)}}-\sum_S{2d^{(S)}(h'^{(S)})^2\over A(h^{(S)})^2}\Big)\alpha(x)\delta(U)\nn\\
&&+\Big(\sum_S {d^{(S)}h'^{(S)} \over Ah^{(S)}}\alpha(x)\Big)\Big({3A'\over A}+\sum_S{d^{(S)}h'^{(S)} \over 2h^{(S)}}\Big)\delta(U)\nn\\
&&-\Big(\sum_S {2\bar\Delta^{(S)} \alpha(x) \over h^{(S)}}\Big)\Big({A'\over A}+\sum_S{d^{(S)}h'^{(S)} \over 2h^{(S)}}\Big)\delta(U)
\ee
The  second term becomes
\be
\Box (\delta R_{UU}) &\dot=&\partial^a  \partial_a  \delta R_{UU}-2\Gamma^b_{aU}   \partial^a \delta R_{Ub}-\Gamma^{ba}_{~~a}  \partial_b \delta R_{UU} -2\Gamma^{ba}_{~~U}  \partial_a \delta R_{bU}-2\delta R_{bU}   \partial^a \Gamma^b_{aU}\nn\\
&\dot=&2g^{UV}\partial_U  \partial_V  \delta R_{UU}+g^{jk}\partial_j  \partial_k \delta R_{UU}-g^{jk}\Gamma^{i}_{jk}  \partial_i \delta R_{UU}-g^{jk}\Gamma^{U}_{jk}  \partial_U \delta R_{UU}\nn\\
&&-2\delta R_{UU}  \,g^{UV} \partial_V \Gamma^U_{UU}\nn\\
&\dot=&\delta(U)\sum_{S,\tilde S} {A\over h^{(S)}h^{(\tilde S)}}~\bar\Delta^{(S)}\bar\Delta^{(\tilde S)}\alpha(x)-{4\over A}\,\Big({A'\over A}+\sum_Sd^{(S)}{h'^{(S)}\over4 h^{(S)}} \Big)^2\alpha(x)\,\delta (U)
\ee
Collecting all we then find the formula 5.
\subsection{Calculation of  $\delta (\nabla^V\nabla_U R)$}
We make following expansion
\be
\delta (\nabla^V\nabla_U R)&=&\delta (g^{Va}\nabla_a\nabla_U R)=(\delta g^{VV})\nabla_V\nabla_U R+g^{VU}\delta (\nabla_U\nabla_U R)\nn\\
&=&(\delta g^{VV})\nabla_V\nabla_U R-g^{VU}(\delta \Gamma_{UU}^a)\nabla_a R+g^{VU}\nabla_U\nabla_U (\delta R)
\ee 
Using the values of $R$ and $\delta R$ in \eqref{R} and \eqref{dR}  above three terms become
\be
&\bullet&\nabla_V\nabla_U R=\partial_V\partial_U R-\Gamma_{VU}^a\partial_a R\,\dot=\, \partial_V \partial_U R\nn\\
&\dot=&{6(A')^2\over A^3}-{4A''\over A^2}-\sum_S\Big({\bar R^{(S)}h'^{(S)}\over (h^{(S)})^2 }+{4d^{(S)}h''^{(S)} \over Ah^{(S)}}-{2d^{(S)}A'h'^{(S)} \over A^2h^{(S)}}+{d^{(S)}(d^{(S)}-7)(h'^{(S)})^2 \over 2A(h^{(S)})^2}\Big)\\
\nn\\
&\bullet&(\delta \Gamma_{UU}^a)\nabla_a R\,\dot=\,(\delta \Gamma_{UU}^V)\nabla_V R+(\delta \Gamma_{UU}^i)\nabla_i R\nn\\
&\dot=&\Big[{6(A')^2\over A^3}-{4A''\over A^2}-\sum_S\Big({\bar R^{(S)}h'^{(S)}\over (h^{(S)})^2 }+{4d^{(S)}h''^{(S)} \over Ah^{(S)}}-{2d^{(S)}A'h'^{(S)} \over A^2h^{(S)}} +{d^{(S)}(d^{(S)}-7)(h'^{(S)})^2 \over 2A(h^{(S)})^2}\Big)\Big]\alpha(x)\delta(U)\nn\\
&&+\sum_S{A\over (h^{(S)})^2}\,(\bar\partial^i\alpha(x))(\bar\partial_i\bar R^{(S)})\,\delta(U)
\\
\nn\\
&\bullet&\nabla_U\nabla_U (\delta R)\,=\,\partial_U\partial_U (\delta R)\,-\Gamma_{UU}^a\partial_a (\delta R)\,\dot=\,\partial_U\partial_U (\delta R)\,\dot=\,0
\ee
where  $\bar\partial^i=\bar g^{ij}\bar\partial_j $.  Collecting all we then  find the formula 6.
\\
\\

\end{document}